

\input harvmac
\input epsf		

\def\MeV{\ifmmode \,\, {\rm MeV} \else MeV \fi}
\def\GeV{\ifmmode \,\, {\rm GeV} \else GeV \fi}
\def\eV {\ifmmode \,\, {\rm eV} \else eV \fi}
\def\myinstitution{\vskip 20pt
   \centerline{\it Enrico Fermi Institute and Department of Physics }
   \centerline{\it University of Chicago, 5640 S. Ellis Ave. }
   \centerline{\it Chicago, IL 60637}
   \vskip .3in
}
\def\myemail{\footnote{$^\dagger$}
        {E-mail: jungman@yukawa.uchicago.edu}}
\def\Fig{Fig.~\the\figno\nFig}
\def\nFig#1{\xdef#1{Fig.~\the\figno}%
\writedef{#1\leftbracket Fig.\noexpand~\the\figno}%
\ifnum\figno=1\immediate\openout\ffile=figs.tmp\fi\chardef\wfile=\ffile%
\immediate\write\ffile{\noexpand\medskip\noexpand\item{Fig.\ \the\figno. }
\reflabeL{#1\hskip.55in}\pctsign}\global\advance\figno by1\findarg}

\def\ifig#1{\noindent\Fig#1 \hbox{: } {\reflabeL{ #1}}}

\def\sbar#1{\kern 0.8pt
        \overline{\kern -0.8pt #1 \kern -0.8pt}
        \kern 0.8pt}  
\Title{\vbox{ \hbox{EFI-92-51}\hbox{hep-ph/9212234} } }
	{P and T Violation From Certain Dimension Eight
	Weinberg Operators }

\def\hisemail{\footnote{$^*$}
        {E-mail: booth@yukawa.uchicago.edu}}

\centerline{ Michael J. Booth\hisemail {\hskip 3pt} {\it and}
	{\hskip 3pt}Gerard Jungman\myemail}
\myinstitution

\vskip .4in
\noindent

Dimension eight operators of the Weinberg type have been shown to
give important
contributions to CP violating phenomena, such as the electric
dipole moment of the neutron. In this note we show how operators
related to these (and expected to occur on equal footing)
can give rise to
time-reversal violating phenomena such as atomic electric
dipole moments. We also estimate the induced
parity violating phenomena such as
small ``wrong'' parity admixtures in atomic states and
find that they are negligible.

\def\jvp#1#2#3#4{#1~{\bf #2}, #3 (#4)}

\def\PRD#1#2#3{\jvp{Phys.~Rev.~D}{#1}{#2}{#3}}
\def\PRL#1#2#3{\jvp{Phys.~Rev.~Lett.}{#1}{#2}{#3}}
\def\PLB#1#2#3{\jvp{Phys. Lett.~B}{\bf #1}{#2}{#3}}
\def\NPB#1#2#3{\jvp{Nucl.~Phys.~B}{#1}{#2}{#3}}
\def\SJNP#1#2#3{\jvp{Sov.~J.~Nucl.~Phys.}{#1}{#2}{#3}}

\lref\thefootnote{ This is also true for operators
	involving three gluons and one photon. These give rise
	to nucleon dipole moments, which will give a non-coherent
	electron-nucleus coupling.}
\lref\EDlims{ S.~Murthy {\it et al}.,  Phys.~Rev.~Lett.
	{\bf 63}, 965 (1989)\semi
	K.~Abdullah {\it et al}.,  Phys.~Rev.~Lett. {\bf 65}, 2347 (1990).}
\lref\bound{ I.~S.~Altarev {\it et al}., \PLB{276}{242}{1992}
	\semi
	\jvp{N.~F.~Ramsey, Ann.~Rev.~Nucl.~Part.~Sci.}{40}{1}{1991}.
	}
\lref\hydcurrent{ how close are we to measuring $10^{-11}$ for hydrogen? }
\lref\weinberg{ S.~Weinberg, Phys.~Rev.~Lett. {\bf 63}, 2333 (1989). }
\lref\Changeight{ D.~Chang, T.~W.~Kephart, W.-Y.~Keung, and T.~C.~Yuan,
	Phys.~Rev.~Lett. {\bf 68}, 439 (1992).}
\lref\Changanom{D.~Chang, W.-Y.~Keung, I.~Phillips and T.~C.~Yuan,
	\PRD{46}{2270}{1992}.}
\lref\Nonrenorm{ R.~Tarrach, \NPB{196}{45}{1982}.}
\lref\mikeanom{ A.~Morozov,
	\SJNP{40}{505}{1984}\semi
	M.~J.~Booth, \PRD{45}{2018}{1992}.}
\lref\mikestandard{ M.~J.~Booth, Enrico Fermi Institute preprint
	EFI-92-13 (1992).}
\lref\QCDstress{ M.~A.~Shifman, A.~I.~Vainshtein and V.~I.~Zakharov
	\PLB{78}{443}{1978}
	.}
\lref\CsTh{ M.~C.~Noecker, B.~P.~Masterson and C.~E.~Weiman,
	\PRL{61}{310}{1988} \semi
	 P.~S.~Drell and E.~D.~Commins, \PRL{53}{968}{1984}
	.}
\lref\hydexp{ R.~R.~Lewis and W.~L.~Williams, Phys.~Lett.
	{\bf 48B}, 111 (1974)\semi
	E.~A.~Hinds and V.~W.~Hughes, Phys.~Lett. {\bf 67B}, 487 (1977).}
\lref\Fischler{W.~Fischler, S.~Paban and S.~Thomas,
	\PLB{289}{373}{1992}.}
\lref\krip{ I.~B.~Khriplovich, {\it Parity Nonconservation in
	Atomic Phenomena} (Gordon and Breach, London 1991)}

\Date{}

Though the discovery of CP violation in the neutral Kaon system
is now a revered part of history,
the origins of CP violation remain mysterious. The observation
of other CP violating phenomena is an important step towards
an understanding of these origins. In particular, the
electric dipole moment of the neutron (EDMN) has inspired much
calculation, as the current experimental limit is now approaching
\def\ecm{ \;e\,{\rm cm}}
$10^{-26} \ecm$ \bound. Quark dipole moments are not the
only source for this effect. Weinberg has introduced a dimension
six operator~\weinberg
\eqn\weinop{ {\cal W}^{(6)} = -{1\over 3} f^{abc} G^a_{\mu\nu}
		G^b_{\nu\sigma} \tilde{G}^c_{\sigma\mu}, }
which can be argued to give an important contribution to the EDMN
in some models. In this paper we will be concerned with the dimension
eight operators
\eqn\GGFf{ {\cal W}^{(8)}_1 = {\kappa_1\alpha_S\alpha\over M^4}
	G^a_{\mu\nu}G^a_{\mu\nu} F_{\sigma\rho}
		\tilde{F}_{\sigma\rho}, }
\eqn\GgFF{ {\cal W}^{(8)}_2 = {\kappa_2\alpha_S\alpha\over M^4}
	G^a_{\mu\nu}\tilde{G}^a_{\mu\nu} F_{\sigma\rho}
		F_{\sigma\rho}. }
The related four gluon operators have been argued to give non-negligible
contributions to CP violating phenomena, such as the EDMN \Changeight.
The anomalous dimensions of these operators have been calculated
\mikeanom, with the interesting conclusion that at least one of the
dimension eight operators is not suppressed by QCD corrections.
The operators ${\cal W}^{(8)}_{1,2}$ will occur naturally whenever the
four gluon operators do.  Because 
$\alpha_S
G^a_{\mu\nu}G^a_{\mu\nu}$ and $\alpha_S G^a_{\mu\nu}\tilde{G}^a_{\mu\nu}$
are unrenormalized to one loop \Nonrenorm, ${\cal W}^{(8)}_{1,2}$ have
vanishing anomalous dimension to that order \Changanom.

In this note we will show that operators of this form can give rise
to observable time-reversal violation in low energy processes, specifically
atomic electric dipole moments. Parity violating effects are estimated
and shown to be small.

{\topinsert
    \epsfbox{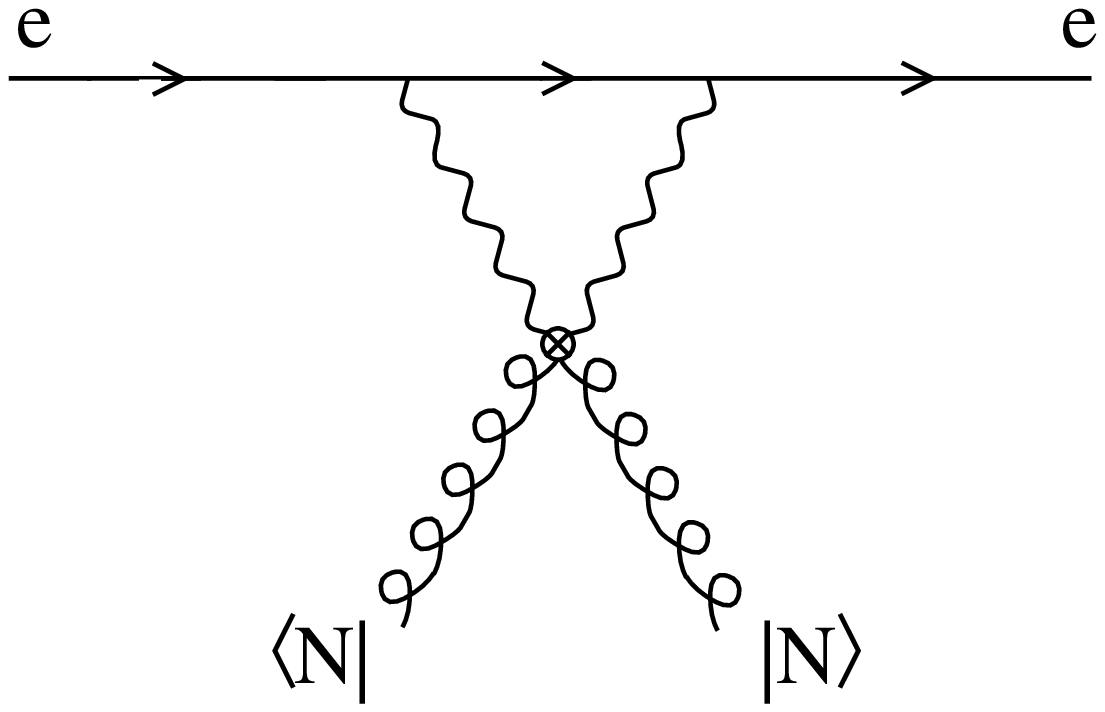}
    \ifig\eeNNfig{Diagram containing the long range contribution of
	${\cal W}^{(8)}_1$ to the parity violating electron-nucleon
	interaction.}
  \endinsert
}

Consider the operator ${\cal W}^{(8)}_1$. Through the diagram
of \eeNNfig,
this operator gives rise to a parity and time-reversal violating
coupling of the electron to two gluons. In order to relate this
to a physical low energy process we must evaluate the gluonic
part of the operator in a hadronic state.
Noting that the gluonic part of ${\cal W}^{(8)}_1$
is equal to the leading part of the
trace anomaly, we calculate
its expectation value in a nucleon state in the standard way,
approximating it by the value at zero momentum transfer \QCDstress.
We write
\eqn\ggNN{ \langle N\mid tr(G_{\mu\nu}G^{\mu\nu}) \mid N\rangle =
	- {8\pi\over b \alpha_S} m_N \langle N\mid \sbar{N} N \mid N\rangle,
}
where $b=11-{2\over 3}n_L$, $n_L$ being the number of quarks
in the effective low energy theory.
Thus ${\cal W}^{(8)}_1$ gives rise to
a parity and time-reversal violating electron nucleon coupling,
\eqn\eeNN{ {\cal O}_{eeNN} =
	{32\over \pi^2}
	{\alpha^2 \kappa_1\over b}
	{m_N m_e \over M^4}
	\;
	f(t)
	\;
	\sbar{e}i\gamma_5 e
	\,
	\sbar{N} N .
}
The function $f(t)$ has the asymptotic behaviours
\eqn\ft{ f(t) = {1\over 16 \pi^2}
	\left\{ \matrix{ {1\over 4} (-t/ m_e^2)\ln( {-t/ m_e^2}),
			\quad & -t/m_e^2 \rightarrow 0^+ \cr
			{3\over 2} \left[\ln({ -t/m_e^2})\right]^2,
			\quad & -t/m_e^2 \rightarrow \infty \cr } \right\}.
}
In the limit $t = (p_e^i - p_e^f)^2 \rightarrow 0$, we have subtracted
simple power law terms which correspond to derivatives of a
delta-function interaction in position space. Such terms are cancelled
by counterterms in the renormalization of the low-energy effective theory.
Note that the operator \eeNN\ gives a coherent nuclear effect,
which is an important enhancement for heavy nuclei. This should be
contrasted with the electron-nucleus coupling induced by nucleon or electron
dipole moments; such a coupling
would be proportional to the spin of the nucleus \thefootnote.

The operator ${\cal W}^{(8)}_2$ involves
a nucleon spin-flip in the non-relativistic limit, which is thus
suppressed by $m_e/m_N$ relative to the operator \eeNN.
It also gives rise to a $\pi^0\gamma\gamma$ vertex
which will induce an electron-nucleon coupling through pion exchange.
However, this operator is suppressed by a factor of $f_\pi /m_N$ relative
to the operator \eeNN.
Thus we consider only the effect of the operator~\eeNN.

In the presence of a static nucleon, the operator of eqn. \eeNN\
gives rise to an effective
perturbation of the electron Dirac equation of the form
\eqn\atompert{ H' = {1\over 2m_e} {i c \,V(r)\,} \vec{\sigma} \cdot \vec{p}, }
where $c$ is the operator coefficient and
$V(r)$ is determined from $f(t)$,
\eqn\Vr{ V(r) = {3\over 16 \pi^3}{1\over r^3}
        \left\{ \matrix{  {(4 r m_e)^{-2}},
		\quad  & r m_e \;\roughly{>}\; 1 \cr
		\ln(r m_e),
		\quad & r m_e \;\roughly{<}\; 1\cr
		}
	\right\}.
}

This perturbation will create a mixing between $s$ and $p$ atomic levels,
similar to the mixing induced by P violating $Z^0$ exchange.
In the case of $Z^0$ exchange, the structure
of the wave function at the nucleus is very important due to the
locality of the potential, and the calculation
of parity-violating mixing effects is difficult. However, for our long-range
potential there is no great subtlety. Of course, treated literally, the
$1\over r^3$ perturbation gives a divergent first Born term, but the
potential is actually mollified by the nuclear size,
$r_N \simeq 1.2\, A^{1/3}\; {\rm fm}$.

In order to estimate the effect of this perturbation on atomic
states, we calculate the induced atomic electric dipole moment
for $\,{}^{55} {\rm Cs}$. The perturbation~\atompert\ gives rise
to a mixing of $\Psi(6s)$ and $\Psi(6p)$ states, thus inducing
an electric dipole moment. For the induced dipole moment we find
\eqn\EDmix{
	\eqalign{
	\mid d_{\rm Cs}\mid &= d_{6s,6p}\;
			\alpha^2 A {\kappa_1\over b}{1\over 2m_e}
			{m_e m_N\over M^4 \Delta E_{\rm 6s-6p}}
			I_O\cr
		&= (2.9\times 10^{-25} \ecm)\, \kappa_1 \,
			\left[{52 \GeV \over M}\right]^4,
	}
}
where $d_{6s,6p}= 2.9\times 10^{-8} \ecm$ is
the measured dipole moment,
$\langle 6s\mid e\vec{r} \mid 6p \rangle$, and $I_O$ is the
matrix element of $ V(r)\, \vec{\sigma} \cdot \vec{p}$.
We have used $b=11$ and $\Delta E_{\rm 6s-6p} = 11000 \;{\rm cm}^{-1}$.
We find that the matrix element is dominated by the contribution from
$r\simeq r_N$, giving
$I_O \simeq 12 \sqrt{35}/\pi ( Z\alpha m_e / 6  )^4 [\ln (r_N m_e)]^2$.
The current limits on such a dipole moment are
$ d_{\rm Cs} < 7.2\times 10^{-24} \ecm$ \EDlims. The implied
limit on the operator coefficient is
$\kappa_1^{1/4} \left( {42\GeV\over M} \right) < 1$. This
limit is even stronger than it first appears since one
typically has $M^4\simeq \Lambda^2 m_Q^2$ where $\Lambda$ is the scale
of the new physics, typically a mass in the Higgs sector, and $m_Q$
is the mass of the heavy quark which is integrated out in the evaluation
of the effective dimension eight operator.
Using $m_Q=m_b$ and $\kappa_1^{1/4}=1$, this translates into
a bound $\Lambda > 350 \GeV$, which is quite restrictive.

We have also estimated the effect of this perturbation on atomic
parity violating phenomena. Note that it is unclear how a practical
measurement of this effect might be accomplished, since it cannot be
measured in polarization experiments in the way that the $Z$ exchange
effect is measured \krip. In any case, the effect is quite small.
We choose to calculate the $2s_{1/2} - 2p_{1/2}$ mixing in hygrogen,
and we find
$| \delta | \simeq 10^{-11}{\kappa_1} \; \left[{1 \GeV\over M}\right]^4$.
An estimate of the mixing induced by $Z$-exchange gives
$| {\delta_{\rm Z}} | \simeq 10^{-11}$, and we see that any
observable effect
is already ruled out by our previous bound.

While completing this work, we became aware of ref. \Fischler, whose
authors, calculating in the context of the minimal supersymmetric
standard model,
consider the contribution of the
operator ${\cal W}^{(8)}_1$ to atomic parity violation.
They use ``naive dimensional analysis'' to
relate
${\cal W}^{(8)}_1$ to the local operator $\sbar{e}i\gamma_5 e\,\sbar{N} N$.
As the authors note, this is may not be a good approximation,
particularly for large Z. Our calculation yields an effect which is
larger than argued in ref. \Fischler.

G.J. was supported by DOE grant DEFG02-90-ER 40560, and
M.B. was supported by DOE grant AC02-80-ER 10587.

\listrefs     
\bye